\documentclass[11pt]{article}
\usepackage[top=0.85in,left=0.75in,right=0.75in, footskip=0.75in]{geometry}
\usepackage[utf8]{inputenc}
\usepackage{comment}
\usepackage{blindtext}
\usepackage{hyperref}
\usepackage{amsmath}
\usepackage{amssymb}
\usepackage{graphicx}
\usepackage{subfig}
\usepackage{amsfonts}
\usepackage{caption}
\usepackage{fancyhdr}
\usepackage{float}
\usepackage[none]{hyphenat}
\usepackage[greek,english]{babel}
\usepackage{afterpage}
\usepackage[utf8]{inputenc}
\usepackage[english]{babel}
\usepackage{color}
\usepackage{longtable}
\usepackage{lscape}
\usepackage{lipsum} % To generate text
\usepackage{array}
\usepackage{flafter}
\usepackage{color, colortbl}
\usepackage[pdftex,dvipsnames]{xcolor}  % Coloured text etc.
\usepackage{amsmath}
\usepackage[colorinlistoftodos,prependcaption]{todonotes}
\usepackage{natbib}
\usepackage[pdftex,dvipsnames]{xcolor}  % Coloured text etc.
\usepackage{amsmath,bm}
\usepackage{xargs}                      % Use more than one optional parameter in a new commands
\usepackage[pdftex,dvipsnames]{xcolor}

\usepackage[colorinlistoftodos]{todonotes}
\newcommandx{\tobedone}[2][1=]{\todo[linecolor=red,backgroundcolor=red!25,bordercolor=red,inline,#1]{#2}}
\newcommandx{\changed}[2][1=]{\todo[linecolor=blue,backgroundcolor=blue!25,bordercolor=blue,inline,#1]{#2}\noindent}
\newcommandx{\thiswillnotshow}[2][1=]{\todo[disable,#1]{#2}}
\newcommandx{\ak}[2][1=]{\todo[linecolor=Orchid,backgroundcolor=Orchid!25,bordercolor=Orchid,inline,#1]{#2}\noindent}
\newcommandx{\df}[2][1=]{\todo[linecolor=BurntOrange,backgroundcolor=BurntOrange!25,bordercolor=BurntOrange,inline,#1]{#2}\noindent}
\newcommandx{\ed}[2][1=]{\todo[linecolor=Aquamarine,backgroundcolor=Aquamarine!25,bordercolor=Aquamarine,inline,#1]{#2}\noindent}
\newcommand{\e}{\mathbf{\hat{\epsilon}}}

\newcommand{\Mm}{\textbf{M}_{1}}

\newcommand{\y}{\textbf{y}}
\newcommand{\kb}[1]{\mathbf{#1}}

\def \X {\kb{X}}
\def \W {\kb{W}}
\def \SO \mathcal{SO}
\def \Or \mathcal{O}
\def \C {\kb{C}}

\def \M {\kb{M}}
\def \Mm {\mathcal{M}}

\def \K {\kb{K}}
\def \B {\kb{B}}
\def \C {\kb{C}}
\def \I {\kb{I}}

\def \1 {\kb{1}}

\title{An application of Saddlepoint Approximation for period detection of stellar light observations}
\author{$\textrm{Efthymia Derezea}^{1}$, $\textrm{Alfred Kume}^{1}$, $\textrm{Dirk Froebrich}^{2}$ }
\begin{document}
\graphicspath{{figures/}} 	
\maketitle
\begin{center}

	\begin{small}
		$1$. \textit{School of Mathematics, Statistics and Actuarial Science, University of Kent, Canterbury CT2 7FS, UK}\\
		$2$. \textit{School of Physical Sciences, University of Kent, Canterbury CT2 7NH, UK}\\
	%	$2$.
	\end{small}	
	\end{center}
\begin{abstract}
One of the main features of interest  in analysing the light curves of stars is the underlying periodic behaviour. The corresponding observations are a complex type of time series with unequally spaced time points. The main tools for analysing these type of data rely on the periodogram-like functions, constructed with a desired feature so that the peaks indicate the presence of a potential period. In this paper, we explore a particular periodogram for the irregularly observed time series data. We identify the potential periods by implementing  the saddlepoint approximation, as a faster and more accurate alternative to the simulation based methods that are currently used. The power analysis of the testing methodology is reported together with applications using light curves from the Hunting Outbursting Young Stars citizen science project.

\end{abstract}

\section{Introduction}\label{s:intr}
%\textbf{correction}
The problem of estimating the periodicity of time series in the presence of irregularly sampled data  appears in many disciplines including Economics (see  \citet{baltagi1999unequally}), Climatology (see \citet{schulz1997spectrum}), Biology (see \citet{heerah2020granger}), or Astronomy (for example for rotation period searches of stars as in \citet{2007prpl.conf..297H}; \citet{2014prpl.conf..433B}). In this paper, we focus on astronomical light curves of stars, whose observations represent brightness measurements over time. 
Current and future large astronomical surveys potentially generate millions of such light curves. Thus, there is a need of accurate, automated, and fast methodologies to determine periods reliably. Depending on the type of star, being able to estimate the period of light curves can provide important information about the star itself (rotation period), its formation, its internal structure and the environment it is in (e.g. structures in accretion disks around young stars).\par 
The sampling of light curves occurs at irregular time points because the data collection which is carried out by ground telescopes can be affected by many factors such as weather, visibility or the telescope's schedule. The data represents the brightness of stars, which are usually quoted in magnitudes $y$. These are determined from the physical flux measurements $f$ as $y = - 2.5 \log{(f/f_0)}$, where $f_0$ is the flux zero point. 
Thus, numerically large magnitude values indicate fainter brightness and vice versa. The times for the data points in the light curves are usually registered as a Julian date. This is a count of days since noon January 1st, 4713 BC.
A typical time series of this type is shown in  Figure \ref{fig:c3314} (Left). 
Note that, each data point has an associated individual measurement accuracy (see the grey bars centered at the observation points there), small values of which reflect small measurement uncertainty. These uncertainties represent one sigma errors and are determined during the data processing and calibration of the astronomical images.

%\begin{figure}
%		\centering
%		\includegraphics[scale=0.45]{figures/lc1v2.png}
%		\includegraphics[scale=0.45]{figures/perlc1.png}
%		\caption{(Left) An example of a light curve as measurement of the brightness over time along with the measurement accuracies (error bars). This is object V1598Cyg measured in the visual filter. (Right) An example of a periodogram, for the light curve shown in the left panel, with F-statistic, based on fitting the sinusoidal model \eqref{sinmod}.}
%		\label{fig:examplelight}
%	\end{figure}

%%%%%%%%%%%%%%%%%%%%%%%%%%%%%%%%%%%%%%%%%%%%%%%%%%%%%
%7896

Let $\y$ denote the vector of $n$ signal observations such that each entry $y_{j}$ is the data arriving in a non systematic way at times $t_{j}$, $j=1,...,n$. We assume the data are generated as 
\begin{equation}\label{general}
y_{j}=g(t_{j};p)+\epsilon_{j} 
\end{equation} 
where $g$ is the assumed true function with period $p$ while  $\epsilon_{j} \sim N(0,\sigma^{2})$ are the measurement errors. If we take the measurement accuracies, which we will denote as $s_{j}$, into account, \eqref{general} changes to $y_{j}/s_{j}=g(t_{j};p)/s_{j}+\epsilon_{j} $ as a weighted regression model.
\par

The main objective is to perform statistical inference about the period $p$ but the challenge is two fold: the first being the  estimation of the period $p$ and the second that of associating some credibility to the claim that the data is indeed periodic. The former problem is addressed by many papers which deal with direct estimation of $p$ but the latter, which is the motivation of the work reported here, is either not paid full attention or done in  computationally inefficient ways.  Depending on the model of interest, a common strategy for the period estimation  is based on the optimal entry of some periodogram function. The period estimation based on linear regression, reported in \citet{thieler2013periodicity},  uses a periodogram function which essentially reflects the regression goodness of fit for each candidate  period and the accepted value of $p$ is that for which the best regression fit is attained. Along the same principle of least square estimation, \citet{hall2000nonparametric} use non-parametric models for identifying the optimal $p$. Obviously, non-parametric approaches make less assumptions about the shape of the light curves, an example of which is the Gaussian process regression (GPR) as seen in \citet{wang2012nonparametric}.

Please note that, when the term ``periodogram" is used, we refer to a broad range of discrete functions defined on a grid of possible period values. Extreme periodogram values (high or low peaks depending on the periodogram chosen) naturally point to potentially valid periods. Figure \ref{fig:per6149} (Right) displays a periodogram (based on the standard F-statistic which we will discuss in Section \ref{s:Hyp}), for a real light curve with period at 2.1763 days as in \citet{froebrich2021survey}. There are many approaches for constructing periodograms depending on the models of interest.  The focus of this paper will be mainly on these non-parametric models and their extensions, such as accounting for correlated residuals and using the additional measurement accuracies in the form of weights. \par
We are particularly interested in addressing here whether a particular peak of the periodogram represents correctly a valid period and  is not just an artifact produced by noisy data. Usually this is addressed by associating p-values and/or performing appropriate hypothesis testing. While this problem is well understood for methods depending on linear models and least squares regression, it is not straightforward how to obtain these p-values when non-parametric regression is used. This is certainly the case for the period detection for irregularly observed time series. For example, in \citet{wang2012nonparametric}, the authors specifically state that a formal test is needed to decide on the validity of a proposed period under such schemes. In this paper, we provide accurate tests to be implemented under such settings, as a faster alternative to the simulation based approaches that are implemented as in \citet{do2009near} for example. A computationally efficient alternative is important given the large amount of data available. Our method improves drastically the calculation time when GPR models are used by performing the calculations in $2\%$ of the time needed for doing the same analysis using Monte Carlo. For more details on time comparisons see Table 4 in Appendix \ref{s:implnum}. In particular, we introduce a general Hypothesis test setting for non-parametric models and specifically Gaussian process regression, providing thus a solution to the problem of period detection for these flexible models. Furthermore, we show how these tests can be adjusted in the presence of correlated noise (red noise), obtaining this way more accurate results while drastically reducing the number of periods which could be falsely identified as valid.\par
Specifically, in this paper we focus on a  hypothesis testing approach for a range of models for periodogram construction, including those derived from non-parametric estimation methods.
In order to  carry out our period estimation for irregularly sampled time series, we will use the generalised F-statistics \eqref{F} to construct the entries of periodogram functions and use the saddlepoint approximation method to evaluate the corresponding p-values for the potential periods. Our method is reported  here for period detection in three sets of models:
\begin{itemize}
	\item 
	 Non-parametric models. This is particularly useful when non-parametric regression (including non-ordinary least squares regression) is used and is shown to be at least as good as the computationally demanding simulation alternatives.
	 \item  Gaussian Process Regression. In particular the leave-one-out cross validation measure used for model fitting leads to the test statistics of a similar quadratic form.
	\item Correlated background noise. These models are extensions of those above while red noise correlation structure is assumed. The saddlepoint method for the resulting quadratic forms has a similar form.
\end{itemize}

%\subsection{Outline}
The outline of the paper is as follows. Current methods and related work are discussed in Section \ref{s:methods}. In Section \ref{s:Hyp} we introduce the general framework of the Hypothesis testing. We see how this framework can be expanded for non-parametric models in Section \ref{s:Hnon}. In Section \ref{s:cvf} we introduce a test based on leave-one-out cross validation error for Gaussian process regression models and in Section \ref{s:hred} we discuss the natural extension of the above mentioned tests under the presence of correlated noise. Specifics on implementation of the Saddlepoint approximation are provided in Section \ref{s:impl}. The tests' performance will be assessed through a detailed power analysis in Section \ref{s:power}, real light curve applications are examined in Section \ref{s:appl}, and finally in Section \ref{s:concl} there is a discussion about future work and conclusions.

\section{Current methods and related work}\label{s:methods}

For the general case of equally spaced time observations the classical periodogram is the estimator of our signal's spectral density: 
\begin{equation} \label{spd}
    P(\omega) =\left|\int g(t)e^{-2\pi i\omega t}dt\right|^{2}
\end{equation}
where $\omega$ is the frequency (it holds that $\omega=1/p$). There are many approaches based on the classical periodogram, see for example  \citet{schuster1898investigation}.

These methods however fail when our observations do not arrive at regular intervals. An illustrative discussion regarding this can be found in \citet{vanderplas2018understanding}, where the author shows how an irregular sampling can lead to a noisy Fourier Transform and thus to noisy estimations of the Spectral density. A popular approach dealing with this problem is that based on the  Lomb-Scargle (LS) periodogram \citep{scargle1982studies} which is a generalized form of the classical periodogram \eqref{spd} but is not affected by irregular time spacing. It can be shown that this method produces identical results to that obtained by fitting a single sinusoidal wave using the ordinary least squares regression while assessing the fit at each period using the squared error, (\citet{lomb1976least}; \citet{vanderplas2018understanding}). The model fitted is basically the same as that of \eqref{sinmod} below but without the intercept term. This model is also studied in \citet{reimann1994frequency} where the author explores the asymptotic behaviour of the maximum likelihood estimator (MLE) of the frequency $\omega$ and proves its consistency based on some regularity conditions regarding the residuals distribution and certain assumptions based on the distribution of time points $t_j$. The author shows that the classic periodogram (moment) estimator of frequency has a larger asymptotic variance than that of the MLE. A very popular model used for the stellar light curves is that based on the sinusoidal functions:
\begin{equation} \label{sinmod}
y_{j}=\beta_{0}+\beta_{1} \sin(2\pi t_{j}/p) +\beta_{2}\cos(2 \pi t_{j}/p)+\epsilon_{j}
\end{equation}
see for example \citet{cumming1999lick} or \citet{vanderplas2015periodograms}. These models are clearly very  special cases of \eqref{general} while there exist many other approaches based on various forms for the periodic function $g$.
For example, \citet{akerlof1994application} fit cubic b-splines using least squares regression in order to estimate the periods of various stars. Another example,  is the use of periodic splines as seen in \citet{oh2004period}, where the authors use cubic splines with a periodic restriction to fit the model by performing robust regression (i.e. Huber's regression) and finally the fit is assessed based on a cross validation approximation. For more general robust regression settings, see for example \citet{katkovnik1998robust}, where the coefficients of a sinusoidal wave are estimated using Huber's M-regression. 
%Another example can be found in \citet{liang2009robust}, where the absolute deviations (L1 norm) regression is used to construct a periodogram in order to estimate the period of gene expression data, showing that these approaches perform better than the L2 norm based methods when the data have outliers. 
A detailed discussion to that end can be found in \citet{thieler2013periodicity} where the authors study, under the same framework, periodograms obtained by combinations of models (such as splines or sinusoidal basis) and different types of regression. In a comprehensive simulation study they compare the performance of the methods for light curves with/without the presence of red noise and show, for example, that robust regression methods are more reliable for data with outliers.\par 
A straightforward application of the simple linear regression method for fitting model (\ref{sinmod}) for various values of $p$ leads to the  construction of some periodogram function whose entries are calculated according to the coefficient of determination $R^{2}$. The valid periods are expected to generate the highest $R^2$ values and hence focusing on the peaks of such periodogram functions is a sensible approach. Additionally, since $R^2$ has a known Beta distribution (see e.g. \citet{schwarzenberg1998distribution}) under the null assumption (of no signal in the data for example) the associated p-values for each potential period reflect the confidence for each proposed period. A similar approach is used in the current literature as in \citet{thieler2013periodicity} where the proposed periods correspond to the highest peaks of the $R^2$ statistics for which the corresponding p-values were low. In that paper the authors calculate the corresponding p-values by fitting to the obtained periodogram a Beta distribution using the Cram\'{e}r-von Mises distance. \par 
Specific interest is shown in period detection under the presence of red noise. The assumed correlation structure of red noise  coincides with that of an AR(1) process if the observations are  collected at equally spaced time observations, see e.g.~\citet{von2001statistical}. In these cases of AR(1) errors, (not of our primary interest here), \citet{benlloch2001quasi} identified valid periodogram peaks against red background noise using Monte Carlo methods. Later some exact tests were developed, see for example \citet{vaughan2005simple}. In the case of unequally spaced time series however most of the corresponding methods used are based on Monte Carlo approaches in order to identify the valid periods in the presence of a red background noise, for example \citet{zhou2002statistical}.\par 

\par 
Other approaches in the literature for period estimation are based on flexible non-parametric models making less assumptions about the shape of the function $g$. A typical example is the use of kernel smoothing regression as studied in great depth in \citet{hall2000nonparametric}, where the estimates for $\hat{g}$ are obtained using the Nadaraya-Watson estimator for a given period and assuming some kernel function. Note that for any given $p$ the parameters of $g$ are immediately obtained based on the minimal sum of squared errors. Hence a grid search on $p$ is sufficient to obtain $\hat{p}$ as well as the parameter estimates of $g$. 
%The parameters for $g$ for a range of candidate periods $p$ are estimated in separate steps, and the proposed optimal period is that which corresponds to the minimal sum of squared errors.
 One could think of the corresponding periodogram function here comprising of values of sums of squared errors for various $p$. 
The authors provide the asymptotic behaviour of their optimal period estimator $\hat{p}$ and show that it is consistent under certain conditions and converges to the true parameter with an $O(n^{-1/3})$ rate. Gaussian process regression for this problem is considered in \citet{wang2012nonparametric}. A prior over functions $g$ is assumed to follow a Multivariate Normal distribution $\kb{g}\sim N(\kb{0},\kb{K})$, where $\kb{K}$ is an $n \times n$ covariance matrix in our case, with entries calculated according to \eqref{kernel} 
\begin{equation} \label{kernel}
    \kb{K}_{jk}=K(t_{j}-t_{k})=A\exp (-2\sin^{2}(\frac{\pi}{p} (t_{j}-t_{k}))/h^{2})\quad  j,k=1,2,\cdots,n 
\end{equation}
and $A$ denoting the amplitude, $p$ the period and $h$ a smoothing parameter. It can be shown that for each candidate period $p$ the corresponding posterior distribution of the fitted function $\bm{g}$ at the time points $t_j$ is Multivariate Normal, with mean and covariance:
	\begin{equation} \label{fitgpr}
	E{(\kb{g})}=\W \y\quad  	\text{Cov}(\kb{g})=\K-\W \K
	\end{equation}
where $\W=\K [\K+\sigma^{2}\kb{I}]^{-1}$ and with $\sigma^{2}$ we denote the variance of our observations.

The parameters are estimated from the marginal likelihood for a given particular period $p$ and similar to pseudo-likelihood methodology the optimal $p$ is chosen to be the maximizer of the resulting marginal likelihood (see more details in Section \ref{s:hred}). Note that if individual measurement accuracies $s_j$ are to be taken into account, a simple update is needed  for using a weighted Gaussian process regression model with,
%\begin{equation}
$\W=\K[\K+\kb{Q}^{-2}\sigma_{n}^{2}\kb{I}]^{-1}$
%\end{equation}
where $\kb{Q}$ is a diagonal matrix with the weights $q_{j}=1/s_{j}$ in its diagonal. 
%\ed{define s here}
The authors in \citet{wang2012nonparametric} claim that their method outperforms that of LS periodogram for data that deviate from the sinusoidal assumption. They also emphasise that a formal statistical test is needed to determine whether the estimated/proposed period can be considered valid. In this paper, we propose a test based on pseudo-likelihood arguments leading to ratios of quadratic forms for normally distributed terms. As the corresponding  distributions are not readily available in closed form, a saddlepoint approximation is implemented and is shown to work well in a range of similar models applied for period estimation.

\section{Hypothesis testing for period detection}\label{s:Hyp}

	As seen in Section \ref{s:intr} the focus of this paper is determining whether a periodogram peak represents a valid period. The common approaches to this problem are  based on running hypothesis tests sequentially and relating them to some periodogram entries. 
	Each periodogram entry could be the value of a chosen goodness of fit test statistic in our case, and we use the corresponding p-value to construct the hypothesis testing whether the given statistics are generated from just noise or period bearing data. In other words, the peaks in the corresponding periodogram point to those candidate periods $p$ which are the most likely to generate the data and their corresponding magnitude of the p-values of the goodness of fit statistics indicate their associated credibility. 
	The idea of sequential hypothesis testing for each candidate period $p$ is summarized as follows:
\begin{align}
H_{0} : & \,\,\Mm_0  \textrm{ model,  no signal in the data}
\nonumber
\\
 \text{vs}&\label{Ho}
\\
H_{1} : & \, \, \Mm_1 \textrm{ model with signal of period }p
\nonumber
\end{align}
%\end{equation}
This strategy is adopted by the astronomy literature as in \citet{schwarzenberg1998distribution}, where the distribution under the assumption of model $\Mm_0$ is derived for empirical periodograms like LS. In many papers the sequential hypothesis testing is based on simulating data according to $H_{0}$ as seen in \citet{do2009near} and then comparing their empirical distribution to the observed values of the periodogram peaks. 

Note that these tests belong to the general family of pseudo-likelihood ratio tests that are based on the relative performance of both models $\Mm_0$ and $\Mm_1$. 
This  approach is in fact similar to that applied to the standard model selection in linear regression for nested models.  
Similar use is seen in time series analysis as shown in  \citet{berenblut1973new} where the authors compare a simple linear model to that with  additional AR(1) errors and the resulting statistic is related to that of \citet{durbin1950testing}.
More specifically, they show that in this context the one sided hypothesis testing is  approximately UMP, see also \citet{paolella2018linear} for an interesting discussion to that end.
In non-parametric regression as seen in \citet{azzalini1993use} a similar test is used for testing the linearity assumption in the data versus non linear models $\Mm_1$ produced by kernel smoothing regression or spline function families.  \par 
	 
Note that similar to  the standard F-tests used in the linear regression, the associated  test statistics for the above mentioned tests are in fact constructed as ratios of residual sums of squares which correspond to quadratic forms of zero mean normal components. The added difficulty including our models of interest is that for each candidate $p$, the terms in the numerator and denominator of those ratios are not necessarily independent. Hence, the standard F-test statistics can not be generally applied. A typical example is that in timeseries for nested ARMA models where standard results for the F-test no longer apply. 
This dependence is not fully addressed in the context of this particular problem. Typically simulation based approaches are implemented to overcome such difficulties.

 In our context, we will perform hypothesis testing as in \eqref{Ho} such that the corresponding model for $\Mm_0$ would  be just a constant, namely no periodic behavior in the data i.e. the light curve is just noise. This is equivalent to the model \eqref{sinmod} reducing to the intercept term $\beta_0$. On the other hand, if there was a periodic signal, the test will suggest as appropriate the alternative fitted models $\Mm_1$ that will contain the periodic terms as in \eqref{sinmod} (or a more general form as in non-parametric regression). As a result, the stronger the evidence of departure form $\Mm_0$ towards $\Mm_1$ for some period $p$, the stronger the claim that this period is present in the signal. Note that a natural goodness of fit measure among various regression models is based on comparing the residual sums of squares $RSS_{0}$ and $RSS_{1}$ for the null and alternative models as:
\begin{equation} \label{F}
F=\frac{(RSS_{0}-RSS_{1})}{RSS_{1} }=  \frac{\y^\top\M_{0}\y-\y^\top\M_{1}\y}{\y^\top\M_{1}\y}
\end{equation}

with matrices $\M_0$ and $\M_1$ depending on the models of interest.  
In order to perform hypothesis testing,  we need to evaluate the distribution of $F$, namely $P(F \leq f)$ under the null hypothesis.  One can easily show that if $\M_{0}=\kb{I}-\kb{1}\kb{1}^\top/n$ is some centering matrix with $\kb{1}$ denoting  the column vector of ones,
\begin{equation} \label{distrFQuad}
P(F>f)=P(\bm{\e}^\top(\kb{I}-(1+f)\M_{1})\bm{\e}>0)
\end{equation}
where $\kb{I}$ is the $n \times n$ identity matrix and $\bm{\e}=\M_{0}\y$.
If we replace $f$ in \eqref{distrFQuad} with a specific observed value of our statistic, e.g $t_{obs}$, the probability $P(F>t_{obs})$ is the p-value for our generalised F-test.\par

\subsection*{Some remarks on the ordinary F-statistics for  linear regression}

In the case of ordinary linear regression \eqref{sinmod} a periodic structure is assumed and the statistic \eqref{F}, if appropriately scaled, coincides with the ordinary F-statistics. The rescaling constant $c=\frac{n-m_{1}}{m_{1}-m_{0}}$
depends on the respective numbers of parameters $m_0$ and $m_1$ for models $\mathcal{M}_0$ and  $\mathcal{M}_1$. 
Namely, under the null assumption of white noise, the statistics \begin{equation}
F \cdot c \sim F(m_{1}-m_{0},n-m_{1})
\label{stF}
\end{equation}
with  the numerator and denominator terms of (\ref{F}) being independent random variables following chi-square distributions with $m_{1}-m_{0}$ and $n-m_{1}$ degrees of freedom. Here, $\M_{0}=\kb{I}-\kb{1}\kb{1}^{T}/n$ is a fixed centering matrix, $\M_{1}=\kb{I}-\X(\X^{T}\X)^{-1}\X^{T}$  with $\X$ being the design matrix of our model depending on observation time points. Note that from the standard calculations of the projection matrix $\kb{P}=\X(\X^\top\X)^{-1}\X^\top$ in regression we can easily see that $\M_{1}=(\I-\kb{P})^\top(\I-\kb{P})=\I-\kb{P}$ and in particular the vector $\kb{1}$ is in the null space of $\M_1$. This implies that $\y^\top\M_{1}\y$ is invariant of the intercept in the model $\Mm_0$ and therefore the statistic \eqref{F} is not affected if the data points $\y$ are centralised, namely replaced by residual vector $\bm{\e}=\M_{0}\y$. Note that under this setting, the eigenvalues of $\M_{1}$ can be either zero or one and that is why \eqref{stF} holds.
The corresponding p-values are immediately available from standard packages like R.

	In situations where more general alternatives are explored for $\Mm_1$, the structure of $\M_1$ does not lead to the standard F-distribution for the statistic \eqref{F} and hence it is not straightforward to obtain its p-values.  Naive Monte Carlo methods are adopted for these cases, and have been extensively used in the relevant literature such as \citet{do2009near} and \citet{halpern2003extreme}, where for each trial period say $p$, many noise curves are simulated and the corresponding p-value is estimated based on the corresponding empirical distribution of the F-statistic (or something equivalent). In fact, we can show that for these cases there is no need to run simulations as the p-values are easily evaluated using the saddlepoint approximation explained in Section \ref{s:impl}. 
\subsection{Generalized F-test for non-parametric periodograms}\label{s:Hnon}

In this section we consider more complicated models $\Mm_1$ to describe our alternative hypothesis and see how the F-statistic as seen in \eqref{F} can be used for non-parametric settings such as kernel or Gaussian process regression (and in general for any linear smoother).
The statistic \eqref{F} has been used in \citet{azzalini1993use} as a linearity test, where a comparison was conducted between a linear and a non-parametric model.

By following the same idea, we compare a constant (as the null model) with a non-parametric model for a given period. Under this setting, while the centering matrix $\M_{0}=\kb{\I}-\kb{1}\kb{1}^\top/n$ remains the same, $\M_{1}=(\kb{I}-\W)^\top(\kb{I}-\W)$ is not a projection matrix,  since $\W$ is a $n \times n$ matrix depending on the observation times and additional model parameters as seen in \eqref{fitgpr} and \eqref{kernel}.

The appropriate period detection test of the optimal periodogram entries requires the evaluation of the distribution of $F$ under the null hypothesis, which due to $\M_1$ can no longer be assumed to follow a standard F distribution. We should note again that this test is also depending on the alternative model through $\M_1$. The probability $P(F>t_{obs})$ of interest, is given by \eqref{distrFQuad} and thus the problem reduces to that of evaluating these values for the distribution of  such Quadratic form of some normally distributed terms.\par 

\par

\subsection{CVF-test for Gaussian process regression}\label{s:cvf}
In the previous section we discussed how the  generalised F-test can be used to create the corresponding periodogram for detecting valid periods when $\Mm_{1}$ is some non-parametric model and the resulting matrix $\M_1$ is not a projection matrix any more. Here we look specifically at the explicit expressions for $\M_1$ when the non-parametric model used is the Gaussian process Regression (GPR)  and  the natural goodness of fit measure based on cross-validation score is used. As seen in Section \ref{s:methods}, in order to estimate the parameters for the GPR model (including the period), we can either maximize the marginal likelihood or minimize the cross-validation error. We can build a statistic/periodogram based on the leave-one-out cross-validation error
\begin{equation} \label{cve}
CVE=\sum_{j=1}^{n}(y_{j}-g_{-j}(t_{j}))^{2}
\end{equation}
 and can use this measure for  comparing the predictive ability of the GPR model at period $p$ with that of the mean $\Mm_0$.
As seen in \citet{williams2006gaussian}, each term in \eqref{cve} can be simplified to
\begin{equation}
y_{j}-g_{-j}(t_{j})=\frac{[(\K+\sigma_{n}^{2}\I)^{-1}y]_{j}}{[(\K+\sigma_{n}^{2}\I)^{-1}]_{jj}}
\end{equation}

and in the vectorized form these entries are shown to be $\y^\top (\K+\sigma_{n}^{2}I)^{-1}\B_{2}$ where $\B_{2}$ is a diagonal matrix,
%\begin{equation}
%\B_{2} = \begin{bmatrix} 
%1/d_{1} & 0 & \dots &0\\
%0 & 1/d_{2} & &\vdots \\
%\vdots &        & \ddots& \\
%0&\dots& &1/d_{n} 
%\end{bmatrix}
%\end{equation}
such that $d_{j}$ is the $j$th diagonal element of the inverse of the matrix $\K+\sigma_{n}^{2}\I$, namely,  $d_{j}=diag((\K+\sigma_{n}^{2})^{-1})_{j}$. Therefore, by denoting as $\B=(\K+\sigma_{n}^{2}\I)^{-1}\B_{2}$, we can write $CVE$ as a quadratic form in $\y$ as

\begin{equation}
CVE= \sum_{j=1}^{n}((\frac{[(\K+\sigma_{n}^{2}\I)^{-1}]_{j}}{d_{j}})\cdot y_{j})^{2}=\y^\top\B \B^\top\y
\end{equation}

In the same spirit as that of \eqref{F},  a similar goodness of fit statistic based on the cross validation errors for the null model and that for a particular $g$ choice using GPR  is given below as:
\begin{equation} \label{cvf}
CVF=\frac{CVE_0-CVE_1}{CVE_1}=\dfrac{\sum_{j=1}^{n}(y_{j}-\bar{y_{-j}})^{2}-\sum_{i=j}^{n}(y_{j}-g_{-j}(t_{j}))^{2}}{\sum_{j=1}^{n}(y_{j}-g_{-j}(t_{j}))^{2}}
\end{equation}
where $\bar{y_{-j}}$ denotes the $j^{th}$ leave-one-out mean. It can be easily shown that the CVF-statistic can be written in the form of \eqref{F} where $\M_{1}=\B\B^\top$ and $\M_0=\M \M^\top$, is a matrix used for the leave one out mean terms, with 1 in the diagonal and $-\frac{1}{n-1}$ everywhere else.
%\begin{equation}
%\M = \begin{bmatrix} 
%1 & -\frac{1}{n-1} & \dots &-\frac{1}{n-1}\\
%-\frac{1}{n-1} & 1 & &\vdots \\
%\vdots &        & \ddots& \\
%-\frac{1}{n-1}&\dots& &1
%\end{bmatrix}
%\end{equation}
The problem again reduces to that of estimating a simple Normal Quadratic form, and Saddlepoint approximation can be implemented in order to obtain fast and accurate results.

\subsection{Testing in the presence of correlated red noise}\label{s:hred}

In this section we consider the situation when a particular correlated noise (red noise) structure is assumed for residual terms $\epsilon$. Let us assume for example that  the correlation structure between any pair of residuals observed at a single time unit  apart is, $Corr(\epsilon_{j},\epsilon_{j-1})=\rho$ for any $j$ and some $\rho\in (-1,1)$. This in turn implies that $Corr(\epsilon_{j},\epsilon_{j-2})=\rho^2$ and in general for any pair of residuals observed at time units $t_j$ and $t_k$,  $Corr(\epsilon_{t_j},\epsilon_{t_{k}})=\rho^{|{t_{j}-t_{k}}|}$.
The red noise terms are distributed as:
\begin{equation} \label{spAR}
   \bm{\epsilon} \sim N(0,\sigma^2 \C_{\rho} )
\quad 
\text{where}
\quad 
\C_{\rho} = \begin{bmatrix} 
1 & \rho^{|{t_{1}-t_{2}}|} &\rho^{|{t_{1}-t_{3}}|} & \dots &\rho^{|{t_{1}-t_{n}}|}\\
& 1 &\rho^{|{t_{1}-t_{2}}|} &\vdots&\rho^{|{t_{2}-t_{n}}|} \\
\vdots &   &     & \ddots& \\
& & &\dots &1 
\end{bmatrix}
%\label{spAR}
\end{equation}
Note that matrix $\C_{\rho}$ takes into account the irregularly sampled nature of the data, such that the correlation between the observations depends on their time distance, the further away two points, the less the correlation assumed for the corresponding residuals. If $\rho=0$, $\C_{\rho}=\I$ then the model reduces to that of the white noise. 
In order to calculate \eqref{F} for various $p$, we need to estimate the model parameters of $\Mm_{1}$. This, for the case of GPR, is achieved by maximizing the marginal likelihood or alternatively by minimizing the leave-one-out cross-validation error. Under the presence of red noise, the marginal likelihood for GPR is:
\begin{equation} \label{loglik}
	\log P(\y|A,h,\sigma^{2},\rho)=-\frac{1}{2}\y^\top(\K+\sigma^{2}\C_{\rho})^{-1}\y-\frac{1}{2}\log |\K+\sigma^{2}\C_{\rho}|-\frac{n}{2}\log 2\pi
	\end{equation}

\par 

For constructing \eqref{F} for the non-parametric models we take $\M_{0}=\C_{\rho}^{1/2}(\kb{I}-\kb{1}\kb{1}^\top/n)\C_{\rho}^{1/2}$ and $\M_{1}=\C_{\rho}^{1/2}(\I-\W)^\top(\I-\W)\C_{\rho}^{1/2}$. Note that matrix $\M_{1}$ can be similarly adjusted accordingly for linear models too. In the case of equally spaced data $\C_{\rho}$ reduces to standard AR(1) correlation structure. In a similar manner as in Section \ref{s:Hnon} and Section \ref{s:cvf}, the p-value $P(F>t_{obs}) $ can be calculated as seen in \eqref{distrFQuad}. The same principles can be applied in order to adjust the CVF-statistic for the correlated noise assumption. See also Appendix \ref{s:redvw} for a comparison between red noise and white noise Gaussian process regression for simulated light curves with additive red noise. 

\section{P-value evaluation for generalised F-test}\label{s:impl}

In this section we will refer to more technical details regarding the implementation of the above mentioned tests.
Note that the corresponding statistics whose p-values need to be generated, are of the form 
\[
F=\frac{\bm{\e}^\top \mathcal{A} \bm{\e} }{\bm{\e}^\top \mathcal{B} \bm{\e}}
\]
where $\mathcal{A}$ and $\mathcal{B}$ are some symmetric matrices of the same dimension $n$ depending on the model and $\bm{\e}$ represent i.i.d normal error components. 
The corresponding p-values for some observed value of statistics $t_{obs}$ is therefore
\[
P(\frac{\bm{\e}^\top \mathcal{A} \bm{\e} }{\bm{\e}^\top \mathcal{B} \bm{\e}}>t_{obs})=P(\bm{\e}^\top (\mathcal{A} -t_{obs}\mathcal{B} )\bm{\e} >0)
\]
Clearly, $\mathcal{A} -\frac{t_{obs}}{c}\mathcal{B} $ is symmetric but not necessarily positive definite, but one could easily see that  
\[
P(\bm{\e}^\top (\mathcal{A} -t_{obs}\mathcal{B} )\bm{\e} >0)=P(X=\sum_{i=1}^l \lambda_i \chi^2_1>0)
\]
Where, $\lambda_i$ are the $l\leq n$ non-zero eigenvalues, not necessarily positive of $(\mathcal{A} -t_{obs}\mathcal{B} )$ and $\chi^2_1$ stands for independent $\chi^2$ random variables with $1$ degree of freedom. Note that the matrices $\mathcal{A}$ and $\mathcal{B} $, when standard linear regression is used, have only two possible eigenvalues: either 0 or 1. This leads to the ratio of two independent chisquare components with $n-m_{1}$ and $m_{1}-m_{0}$ degrees of freedom respectively which reduces to the standard F distribution (see \citet{butler2007saddlepoint}, p. 378). As the distribution of $X$ is not generally known in the closed form, see \citet{johnson1995continuous}, we need to evaluate these probabilities numerically. There are many methods in the literature for approximating such probabilities, for example numerical integration \citet{imhof1961computing} or using the method of matching moments. The moment generating function of the corresponding convolution $X$ is known however and rather than performing the corresponding numerical Laplace inversion, a practically convenient method that we adopt here is that based on the saddlepoint approximation as it is numerically efficient for the accuracy that we need to operate. Moreover, the saddlepoint approximation has been successfully implemented in many timeseries applications e.g. \citet{paolella2018linear} and more recently in non euclidean statistics data e.g. \citet{kume2013saddlepoint}. We adopt the methodology that is suggested in \citet{kuonen1999miscellanea} as our hypothesis testing methodology  is also similar in nature.

\subsection{Saddlepoint approximation }

	Saddlepoint approximation is an approximation method for evaluating a density or a cumulative distribution function given the analytical expression of its moment generating  function or the cummulant generating function $\mathcal{K}(\cdot)$ and its derivatives $\mathcal{K}'$ and $\mathcal{K}''$, see \citet{butler2007saddlepoint} for more details. In fact, our focus here is on  the CDFs, namely $F(x)=P(X>x)$. Two popular approximations are available for these survival probabilities, as suggested in \citet{lugannani1980saddle} and \citet{barndorff1990approximate}. In particular, the approximation of \citet{barndorff1990approximate}  for $\hat{F}(x)$ is derived as
\begin{equation}
\hat{F}(x)=1-\Phi\{w+\frac{1}{w}\log (\frac{u}{w})\}
\end{equation}
where $\Phi$ is the standard normal distribution, and $w$ and $u$ are given by,
$w=sgn(\hat{s})[2(\hat{s}-\mathcal{K}(\hat{s}))]^{1/2}$,  $u=\hat{s}(\mathcal{K}''(\hat{s}))^{1/2}$. The saddlepoint solution is $\hat{s}$  such that 
\begin{equation} \label{spe}
\mathcal{K}'(\hat{s})=x
\end{equation} 

In our case, we consider the quadratic forms of zero mean normal variables. It can be shown that for some independent random variables, $Z_i\sim N(0,1)$, the corresponding cummulant generating function of the quadratic form $\sum \lambda_i Z_i^2$   is $\mathcal{K}(s)=-\frac{1}{2}\sum \log (1-2s\lambda_{i})$ with, $\mathcal{K}'(s)=\sum \frac{\lambda_{i}}{1-2s\lambda_{i}}$ and $\mathcal{K}''(s)=\sum \frac{2\lambda^2_{i}}{(1-2s\lambda_{i})}$. As can be seen from the expression for $\mathcal{K}'$, at $s=1/(2\lambda_i)$ this function is unbounded and therefore the solution of $\hat{s}$ should be carefully chosen. In particular, we have found that the best strategy here is to look for the solution which is the nearest to $0$. This implies that the solution of the saddlepoint equation (\ref{spe}) can be also negative. Some examples regarding the accuracy of saddlepoint approximation for our particular problem can be found in Appendix \ref{s:implnum}.

%%%%%%%%%%%%%%%%%%%%%%%%%%%%%%%%%%%%%%%%%%%%%%%%

\section{Power analysis of the tests}\label{s:power}

	In this section we examine the performance of the proposed tests by estimating their power using simulations. The power is the the probability to identify a period as valid when the period is correct, and it is usually denoted as $\gamma$. In other words $\gamma=1-P(\textrm{Do not  reject }H_{0}|H_{1}\textrm{ true})$. The higher the value of the power the better the performance of the test.\par 
	\subsection{Example 1: Weighted linear regression}
\begin{figure}
		\centering
		\includegraphics[scale=0.46]{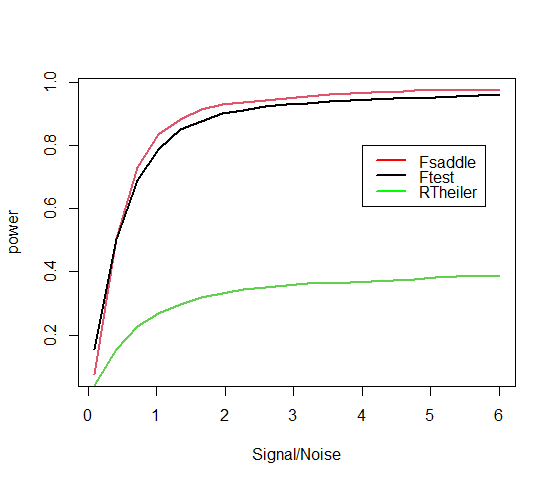}
		\includegraphics[scale=0.46]{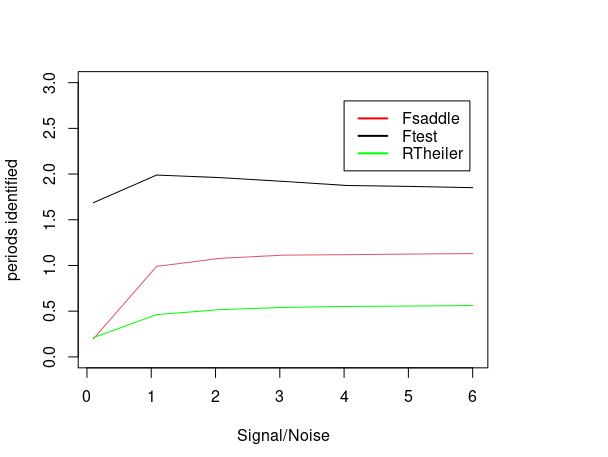}
		\caption{(Left) Comparison of the power between the F-test with Saddlepoint approximation, standard F-test and Theiler's approach, for different signal to noise ratios. (Right) Comparison between the average number of periods identified as correct for the three tests mentioned. For each value of SNR 1000 curves with 200 data points were generated. The significance level was set to $\alpha=1-0.998$.}
		\label{fig:powerweights}
	\end{figure}

We estimate $\gamma$ using simulated light curves with known periodicity at 2.4 days. We mentioned in Section \ref{s:Hyp} that in many cases the distribution of the F-statistic \eqref{F} no longer follows a standard F distribution. One of these situations is when weighted regression is used. As a first example, we compare the power for the generalized F-test to that of the standard F-test and to the approach described in \citet{thieler2013periodicity} when weighted linear regression is used. For this example $\alpha= 1-0.998$. We simulate 1000 sinusoidal light curves with known period consisting of 200 data points for different signal to noise ratios (SNR), $SNR=var(y)/var(\epsilon)$. 
These light curves will be generated using the package \textit{RobPer} \citep{thieler2016robper}. The results are shown in Figure \ref{fig:powerweights} (Left) for the significance level $\alpha=1-0.998$, note that this level is chosen as the multiple tests correction for $\alpha=0.05$ and 30 repetitions according to \citet{vsidak1967rectangular}. We notice that  the generalised F-test outperforms the other two while closely followed by the standard F-test. Of course the larger the SNR, the larger the power, meaning that the probability of correct inference, which depends on the estimated p-values, gets larger as the SNR increases.\par
In order to include the testing approach introduced in \citet{thieler2013periodicity}, which we denote here as ``RThieler", for each generated light curve we calculated the relevant periodgram for 196 trial periods ranging from 0.5 to 20 and used the \textit{RobPer} package to calculate the relevant critical value. To obtain a clearer view on the test's performance we further calculate the average number of periods identified as correct, for different signal to noise ratios. In this example we are searching for periods from 0.5 to 10 days with 1 decimal accuracy. Note that the simulated light curves were generated with only one actual period, and thus extra periods detected are false alarms. The results can be seen in Figure \ref{fig:powerweights} (Right), in this plot the number of periods detected should be 1 and thus deviations from that are an indication of poor performance. We can see that the generalized F-test clearly outperforms the other two, with the standard F-test overestimating the number of correct periods and the RThieler test underestimating it. Overall, from this example we can conclude that using the generalized F-test when weighted regression is used can improve the quality of our results compared to the other two methods.

%%%%%%%%%%%%%%%%%%%%%%%%%%%%%%%%%%%%%%%%%%%%%%%%%%%%%

%%%%%%%%%%%%%%%%%%%%%%%%%%%%%%%%%%%%%%%%%%%%%%%%%%%%%

%\begin{table}

%	\begin{center}
%		\begin{tabular}{|c|c|c|}
			
%			\hline
%			\multicolumn{3}{|c|}{power comparison }\\
%			\hline
%			& \textbf{white noise}  &\textbf{red noise} \\
%			%\textbf{C.V} & \textbf{Beta } &\textbf{Bootstrap } &\textbf{Beta }  &\textbf{Bootstrap}\\
%			$F*$  &  0.886&0.866\\
%			%sinusoidal sig& 3 &2   \\
%			$CVF*$& 0.911 &0.882 \\
%			Monte Carlo& 0.902 &0.858   \\
			
%			\hline
			
%		\end{tabular}
%	\end{center}
%	\caption{Comparison of the power of the proposed tests for white and red background noise scenarios, for a light curve with amplitude 1 and period 2.4 and variance 1. }
%	\label{tab:power1}
%\end{table}
\subsection{Example 2: Gaussian process regression}\label{s:gprpow}

\begin{figure}[h]
		\centering
		\includegraphics[scale=0.46]{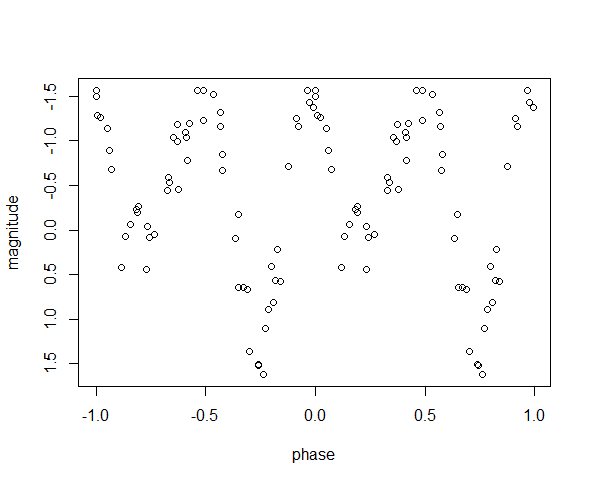}
		\includegraphics[scale=0.46]{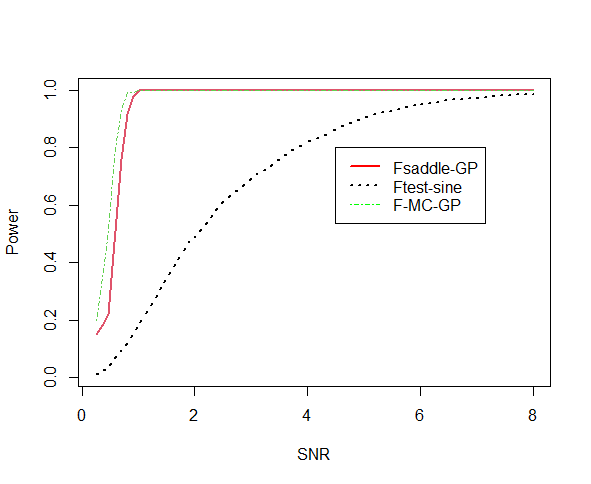}
		\caption{(Left) A phased light curve generated from the GPR model with period 5.2 days with SNR=6. This is a typical example of the shape of the light curves generated. (Right) The estimated power of the GPR generalized F-test of light curves generated for different SNR (red line). the green line shows the power when the same analysis is perform using Monte Carlo and the dashed line shows the power of the standard F-test for a sinusoidal based fitting.}
		\label{fig:exampleimage2}
	\end{figure}

In Section \ref{s:Hnon} and Section \ref{s:cvf} we discussed that the generalized F and CVF tests can be readily used for more complex models such as Gaussian process regression. In this section we simulate artificial light curves from a GPR model, fit a Gaussian process regression and perform our tests in order to estimate their power. For this example we borrow the sampling of a randomly selected real light curve (object 3314$\_$I from \citet{froebrich2021survey}). We generate our periodic signal from a Gaussian prior using a periodic kernel as in \eqref{kernel} with period at 5.2 days. A typical example of the shape of our simulated data can be seen in Figure \ref{fig:exampleimage2} (Left). We generate light curves for different signal to noise ratios, ranging from 0.01 to 8. The noise is generated from a zero-mean Normal distribution. We estimate the power for each different value of SNR based on 1000 repetitions. In Figure \ref{fig:exampleimage2} (Right) we see the power of the generalized F-test at a $1-\alpha$ =99$\%$ significance level. We notice that the test performs very well with its power being estimated larger than 0.7 for even SNRs as small as 0.5. For a SNR larger or equal to 1 the power is constantly estimated close to 1. \par 
To our knowledge there doesn't exist another test under the GPR setting to pose as a comparison. We include in our plot the estimated power of the standard F-test using a sinusoidal model, for the same simulated data, as a reference, acknowledging that the comparison of these tests is not fair. The generalized F-test performs a lot better than the standard F-test in this example as expected. We performed the same analysis for the CVF-test too and obtained very similar results to those obtained from the generalized F-test. In Table 1 we can see a comparison of the power of generalized F and CVF tests for the same simulated data and SNR fixed to 0.65. The power is calculated for different significance levels. Both tests perform very well with their power estimated to be larger than 0.84 for a significance level of $1-\alpha=99.5\%$ for example. Of course the larger the $1-\alpha$ significance level the smaller the estimated power. We should note that the significance levels were chosen as multiple testing corrections according to \citet{vsidak1967rectangular} and correspond to different numbers of tests (e.g. 1, 10, 100, 1000) that could be conducted. We notice that the tests behave similarly with the CVF-test having a slightly larger power in this particular example. Finally we also calculated the probability of not rejecting the null hypothesis when the null is true by simulating purely noise data. In all cases the probability was larger or equal to 0.98 for our proposed tests and the same holds for when a linear model and standard F-test is applied.

\begin{table}
\caption{This table shows  the estimated power of the generalized F and CVF tests for different significance levels.}
\begin{centering}
\begin{tabular}{lrr}
 
\textbf{sig. level}&\textbf{$\mathbf{F}$ GPR }&\textbf{CVF GPR}\\
\hline
$0.95^{1/1}=0.95$& 0.968 &0.970  \\
$0.95^{1/10}=0.9948838$& 0.842 &0.848 \\
$0.95^{1/100}=0.9994872$& 0.616 &0.612\\
$0.95^{1/1000}=0.9999487$& 0.352 &0.354 
\\

\end{tabular}\hfill
\end{centering}
\label{tab:simGP2}
\end{table}

\section{Application to real light curves}\label{s:appl}

In this section we apply the methodology from the previous sections to real light curves. They are obtained as part of the Hunting Outbursting Young Stars Citizen Science Project \citep{2018MNRAS.478.5091F}. This project combines observational data from professional, university and amateur observatories to construct long-term light curves of young stars. By its very nature the project hence creates in-homogeneously sampled light curves. The example objects investigated here are all situated in the Pelican Nebula, a vast star forming region in Cygnus. Periodic light curves of young stars can be used to measure rotation periods of the stars, as well as to track the evolution of properties of surface features on them.

\par 

\subsection{Example 1} 

\begin{figure}[ht]
		\centering
		\includegraphics[scale=0.46]{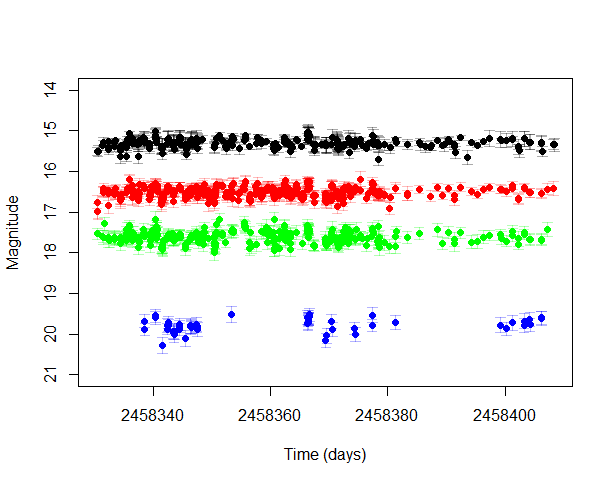}
		\includegraphics[scale=0.46]{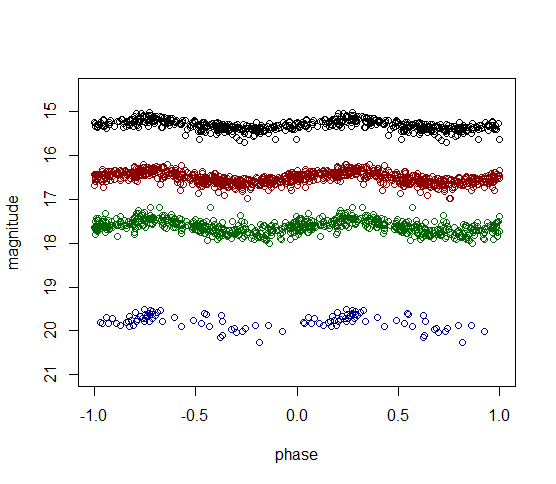}
		\caption{(Left) The light curve number 6149. Each color represents a different filter. Black is for Infrared, red for Red filter, blue for Blue filter and green for Visual. The error bars represent the measurement accuracies. (Right) The light curve number 6149 folded in period 2.1763 days for different filters.}
		\label{fig:c6149}
	\end{figure}

We first investigate object number 6149 from \citet{froebrich2021survey}. This object belongs to the population of young stars in the Pelican Nebula and shows a clear periodic behaviour in the four filters B, V, R, and I. The period of the variations has been determined as 2.1763\,d, from the median of the periods in the individual filters \citep{froebrich2021survey}. See Figure \ref{fig:c6149} where the light curve is plotted for the different filters against time and folded in period 2.1763 days.

\begin{figure}[h!]
		\centering
		\includegraphics[scale=0.42]{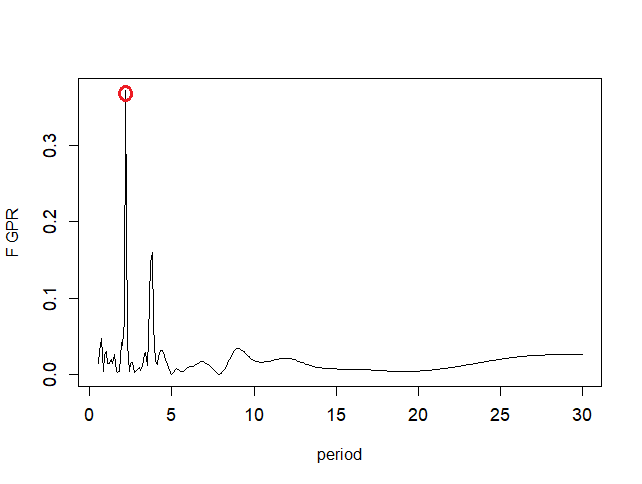}
		\includegraphics[scale=0.42]{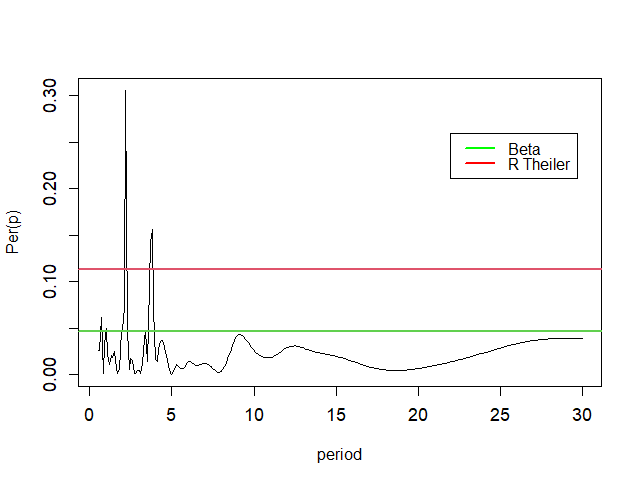}
		\caption{(Left) The weighted GPR $F$ periodogram for light curve 6149 measured in the visual filter. With the red circle we mark the period identified as important according to the generalized F-test. (Right) The weighted least squares sine periodogram for the same light curve. The lines represent the critical values according to the Standard F-test and the ``RTheiler" approach, peaks above the lines are considered as important periods.}
		\label{fig:per6149}
	\end{figure}

%%%%%%%%%%%%%%%%%%%%%%%%%%%%%%%%%%%%%%%%%%%%%%%%%%%%
We first analyze the light curve independently for each filter, by fitting a weighted Gaussian process model using the periodic kernel and assuming independent residuals. In principle, light curves measured in different filters for the same star should exhibit the same periodic behaviour, so if a method identifies the same period for all filters it is a good sign that the correct period is detected. The periodogram is obtained using F and CVF statistics, see Figure \ref{fig:per6149} (Left). For this example we search for periods between 0.5 and 30 days with a very rough grid of 1 decimal accuracy. We then apply sequentially the generalized F and CVF tests and identify in most cases the period at 2.2 days as the most important real period. This is the period that corresponds to the maximum F-statistic value and a significant p-value. We notice however, that for some filters the tests identify more than one period as valid and in some cases as in filter R the tests fail to identify any periods. We further run the same analysis but this time assuming some correlation structure for the residuals as described in Section \ref{s:hred}. We see that with these adjustments the generalized F and CVF tests return as valid only one period at 2.2 days also. It is worth noticing that for filter R the tests identify also the period at 2.2 days.
\par Next, for comparison, we analyze the lightcurve independently for each filter again, by fitting a sinusoidal wave using weighted least squares regression. We obtain the periodogram based on the $R^{2}$ statistic seen in Figure \ref{fig:per6149} (Right). We then proceed by performing the standard F test and Beta distribution based test as seen in \citet{thieler2013periodicity}. The results for all methods and filters are summarized in Table 2. We see that all methods in most cases identify the most important period as 2.2. The standard F test seems to be overestimating the number of valid periods. All methods seem to find as valid other periods than 2.2, with some exceptions for the GPR based tests.

%%%%%%%%%%%%%%%%%%%%%%%%%%%%%%%%%%%
%The dash represents the case that a test did not identify any period as valid.
\begin{table}
%\captionsetup
\caption{The period identified as most important from the various tests, for different filters. The numbers in the parenthesis denote the number of other periods identified as important from the tests. The dash represents the case that a test did not identify any period as valid.}
\centering
\resizebox{\textwidth}{!}{%
\begin{tabular}{lrrrrrr}
\textbf{Filter} & \textbf{$\mathbf{F}$ GPR }  &\textbf{CVF GPR} &\textbf{F GPR red}& \textbf{CVF GPR red}&\textbf{F sine }&\textbf{Thieler sine}\\
\hline
B& 2.2 (2 extra) &2.2 (2 extra)  &2.2 (0 extra)&2.2 (0 extra)&2.2 (2 extra)&- \\
R& - &- &2.2 (0 extra)&2.2 (0 extra) &2.2 (3 extra)&2.2 (1 extra) \\
I& 2.2 (1 extra) &2.2 (1 extra)&2.2 (1 extra)&2.2 (1 extra)  &2.2 (1 extra)&2.2 (1 extra) \\
V& 2.2 (0 extra) &2.2 (0 extra) & 4.3 (1 extra) &4.3 (1 extra) &2.2 (3 extra)&2.2 (1 extra) 
\\
\end{tabular}%
}
\hfill
\label{tab:appl2}
\end{table}

%%%%%%%%%%%%%%%%%%%%%%%%%%%%%%%%%%%
%\begin{table}
%\captionsetup{width=0.8\linewidth}
%\caption{The period identified as most important from the various tests, for different filters. The numbers in the parenthesis denote the number of other periods identified as important from the tests. The dash represents the case that a test did not identify any period as valid.}
%\begin{centering}
% \noindent\resizebox{\linewidth}{!}{%
%\begin{tabular}{lrrrrrr}
% \multicolumn{1}{c}{} &
%      \multicolumn{6}{c}{Method} \\ 
%\textbf{Filter} & \textbf{$\mathbf{F}$ GPR }  &\textbf{CVF GPR} &\textbf{F GPR red}& \textbf{CVF GPR red}&\textbf{F sine }&\textbf{Thieler sine}\\
%\hline
%B& 2.2 (2 extra) &2.2 (2 extra)  &2.2 (0 extra)&2.2 (0 extra)&2.2 (2 extra)&- \\
%R& - &- &2.2 (0 extra)&2.2 (0 extra) &2.2 (3 extra)&2.2 (1 extra) \\
%I& 2.2 (1 extra) &2.2 (1 extra)&2.2 (1 extra)&2.2 (1 extra)  &2.2 (1 extra)&2.2 (1 extra) \\
%V& 2.2 (0 extra) &2.2 (0 extra) & 4.3 (1 extra) &4.3 (1 extra) &2.2 (3 extra)&2.2 (1 extra) 
%\\
%\end{tabular}
%}\hfill
%\end{centering}
%\label{tab:appl2}
%}
%\end{table}

%%%%%%%%%%%%%%%%%%%%%%%%%%%%%%%%%%%%%%%%%%%%%%%%%%%%%

\subsection{Example 2}
\begin{figure}[h!]
		\centering
		\includegraphics[scale=0.45]{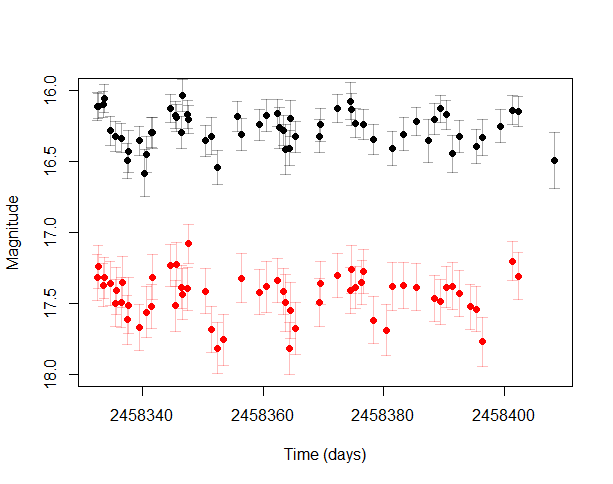}
		\includegraphics[scale=0.45]{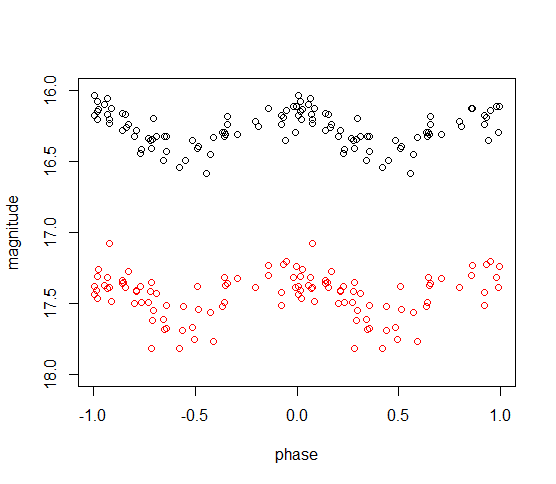}
		\caption{(Left) The light curve number 3314. Black is for Infrared filter and red for Red filter. The error bars represent the measurement accuracies. (Right) The light curve number 3314 folded in period 13.8783 days for different filters.}
		\label{fig:c3314}
	\end{figure}

As a second example, we analyze object 3314, measured in filter I and R with a studied period at 13.8783 days, as seen in \citet{froebrich2021survey}. See Figure \ref{fig:c3314} (Left) where the light curve is plotted for both filters and Figure \ref{fig:c3314} (Right) where the light curve is folded in a 13.8783 day period. Similarly to example 1 we apply all previous methods to both filters. The results are summarized in Table 3. We see that for this example all methods behaved in the same way. They all identified only one important period at 13.9 days. In this example the red noise model was applied too returning the same results and was excluded from the table for simplicity.

%%%%%%%%%%%%%%%%%%%%%%%%%%%%%%%%%%%%%%%%%%%%%%%%%%%%%

%Similarly to example 1 we apply all previous methods to both filters. The results are summarized in Table 3. We see that for this particular example all methods behaved in the same way. They all identified only one important period at 13.9 days. In this example the red noise model was applied too returning the same results and was excluded from the table for simplicity.
%%%%%%%%%%%%%%%%%%%%%%%%%%%%%%%%%%%%%%%%%%%%%%%%%%%%

\begin{table}
%\captionsetup{width=0.8\linewidth}
\caption{The period identified as most important from the tests in Example 2, for different R and I. In this case no other period was identified as important from the tests (0 extra periods).}
\begin{centering}
\begin{tabular}{lrrrr}
% \multicolumn{1}{c}{} &
%      \multicolumn{4}{c}{Method} \\ 
\textbf{Filter} & \textbf{F GPR }  &\textbf{CVF GPR} & \textbf{F sine }&\textbf{Thieler sine}\\
\hline
R& 13.9 (0 extra) &13.9 (0 extra)  &13.9 (0 extra)&13.9 (0 extra) \\
I& 13.9 (0 extra) &13.9 (0 extra)  &13.9 (0 extra)&13.9 (0 extra) \\

\\

\end{tabular}\hfill
\end{centering}
\label{tab:appl1}
\end{table}

%%%%%%%%%%%%%%%%%%%%%%%%%%%%%%%%%%%%%%%%%%%%%%%%%%%%%%%%%%%%%%%%%%
\subsection{Implementation notes}
For the examples in this section we used a rough period grid of one decimal accuracy. This grid is chosen for computational simplicity and it also shows us how methods perform under relatively crude search schemes. In general it is advised to follow the two stage period search method seen in \citet{reimann1994frequency} and described in Appendix \ref{s:redvw}. Another thing to note is that we selected our period grid homogeneously in the range between 0.5 and 30 days. Here we had prior knowledge as to where the potential periods were, since they were studied in \citet{froebrich2021survey}, and so this approach is sensible as the results also showed. In general applications however, when no prior information is available, it is preferable to select periods in-homogeneously, by building a grid based on the equivalent frequency range instead and ensure that short periods won't be missed.

\section{Conclusions and future work}\label{s:concl}
In this paper we provided tests for period detection under non-parametric linear smoother settings, especially Gaussian process regression. The generalized F-statistic is easily and readily adjustable for a range of models. For example it works better than alternative tests when Weighted Least Squares regression is used and the sample size is relatively large. It is also easy to use under ARMA model setting providing more accurate results than asymptotic methods for small sample sizes. The CVF-test statistic based on leave-one-out cross validation for Gaussian process regression behaves similarly to the F. Both statistics can be adjusted for Weighted Gaussian process regression scenarios (when measurement accuracies are available), and also for deviations from white noise. Our simulation results show that adjusting for red noise (when present) leads to more periods correctly identified and smaller number of falsely identified periods.
The power of both tests when Gaussian process regression is used is quite high for reasonable signal to noise ratios (e.g. for data with $SNR>0.6$). \par 
These tests are flexible but there are situations where they cannot be used (e.g L1 regression settings). We have some preliminary results for a test based on bootstrap that can be used under any setting that will be reported as part of a future work. Finally, the methods could be adjusted to take in to account simultaneous measurements from multiple filters. Further research is needed for situations where the data exhibit semi-periodic behaviour. Asymptotic results on the estimator of period under the Gaussian process regression settings could provide useful insights about efficiently calculating periodograms.
\section*{Acknowledgements}
The authors would like to thank the anonymous referees for improving this paper with their comments.
\section*{Data availability statement}
The data is available upon request from the authors.
\section*{Funding statement}
This work was funded by the University of Kent Vice chancellor's scholarship.
%\bibliographystyle{rss}
%\bibliography{example}
\bibliography{sample}
\appendix
%\end{comment}

%\newpage
%\appendix
\section{Appendix}
 
\subsection{Exploring the Saddlepoint approximation accuracy} \label{s:implnum}
In this subsection we will see some numerical examples in order to understand further the behaviour of the Saddlepoint approximation for the statistics proposed in this paper. We will start by approximating the cumulative distribution function (CDF) of the generalized F-statistic under the null hypothesis when the alternative is a periodic Gaussian process model.
	\begin{table}
%% Caption MUST come immediately after \begin{table}
\caption{\label{tab:times}Time in seconds for 50 repetitions of the different methods.}
\centering
\begin{tabular}{lrrrr}
 & Saddlepoint & Imhof & Monte Carlo & Exact \\\hline
GPR &  16.30& 17.56& 774.90 &\textendash\\
OLS& 0.79& 1.24 &4.10 & 0.02   \\
%Imhof&1.234778\\
%Monte Carlo& 4.098558
\end{tabular}
\end{table}

\begin{figure}
		\centering
		\includegraphics[scale=0.47]{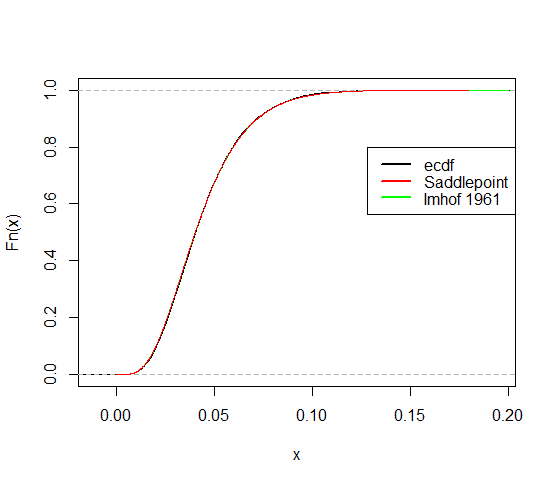}
		\includegraphics[scale=0.47]{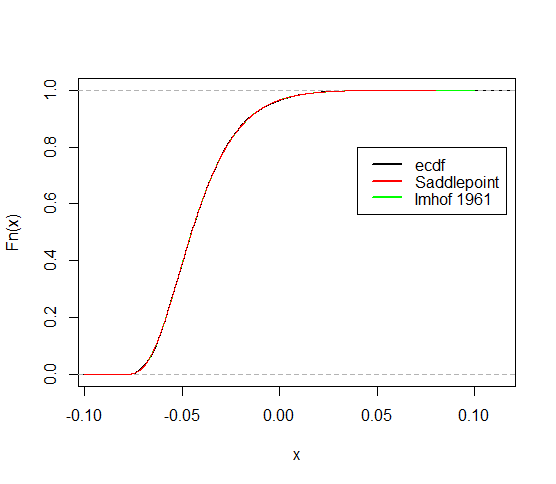}
		\caption{(Left) An example of the ECDF and the saddlepoint approximation of the corresponding generalized F-statistic based on the RSS of White Noise and GPR models. The green line that agrees with the saddlepoint approximation is the approximation based on the numerical integration method. (Right) The same comparison of methods for the CVF-statistic of GPR.}
		\label{fig:Fstarmc}
	\end{figure}

In order to examine the performance of Saddlepoint approximation, we simulate 10000 values of the generalized F-statistic under the null hypothesis assumption and calculate its empirical cumulative distribution function (ECDF), which we compare with the CDF obtained by using Saddlepoint, see Figure \ref{fig:Fstarmc} (Left). Saddlepoint approximation gives results that are very close to the empirical CDF of the statistic. In addition, we plot the CDF approximated using numerical integration according to \citet{imhof1961computing} getting almost the same results with the other approaches. The only difference is that Saddlepoint approximation is faster, see more details at Table \ref{tab:times} for a time comparison between the methods. Please note that whenever we report ECDF values below we mean the empirical CDF of 10000 replications. Similar results have been produced for the saddlepoint approximation of the CVF-statistic. In Figure \ref{fig:Fstarmc} (Right) we see that the saddlepoint approximation of $CVF$ under the null when GPR is used as the alternative also works well. \par

%%%%%%%%%%%%%%%%%%%%%%%%%%%%%%%%%%%%%%%%%%%%%%%%%%%%%%%%%%%%%%%%\caption{\label{tab:widgets}An example table.}
%\centering
%	\begin{table}
%	\caption{\label{tab:times}The time in seconds for the methods used to determine the valid periods for one light curve and a periodogram of 50 trial periods. The periodogram was based on fitting a sine wave using least squares regression. }
%		\centering
%			\begin{tabular}{|c|c|}
		%	\hline
%				 \multicolumn{1}{|c|}{\textbf{Method}} & \multicolumn{1}{|c|}{\textbf{Time in seconds}}\\\hline
%				Theoretic &  0.01681399\\
%				Saddlepoint& 0.789396    \\
%				Imhof&1.234778\\
%				Monte Carlo& 4.098558   
			%	\hline
%				\end{tabular}
%		\end{center}
%		\caption{The time in seconds for the methods used to determine the valid periods for one light curve and a periodogram of 50 trial periods. The periodogram was based on fitting a sine wave using least squares regression. }
		%\label{table}
%		\label{tab:times}
%	\end{table}
	%%%%%%%%%%%%%%%%%%%%%%%%%%%%%%%%%%%%%%%%%%%%%%%%%%%%%%%%%%%%%%%%%

	%%%%%%%%%%%%%%%%%%%%%%%%%%%%%%%%%%%%%%%%%%%%%%%%%%%%%%%%

\begin{figure}[ht]
		\centering
		\includegraphics[scale=0.45]{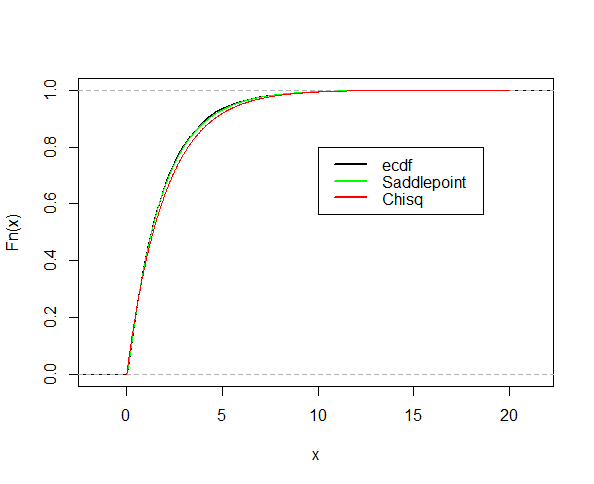}
		\includegraphics[scale=0.45]{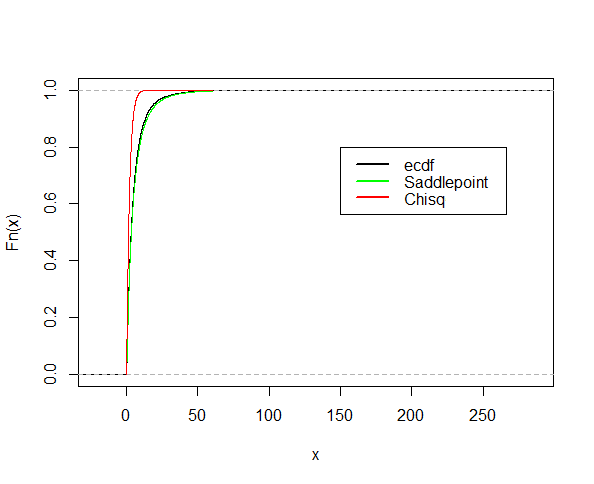}
		\caption{(Left) A comparison of the CDF of the F-statistic under the null of correlated noise for a light curve of 200 observations. The black line shows the empirical CDF, the green line corresponds to Saddlepoint approximation and the red line is the asymptotic Chisquare distribution. (Right) A comparison of the CDF of F-statistic under the null of correlated noise for a light curve of 200 observations. Note that when the sample size is very small the asymptotic result no longer holds but the saddlepoint method approximated the CDF quite accurately.}
		\label{fig:spsamplesize}
	\end{figure}

The hypothesis testing proposed can be also used to model correlated background noise and in many cases can perform better than standard asymptotic tests. In this example we will compare the approximations of the CDF of the F-statistic under the presence of AR(1) type noise and when the alternative is fitting a sinusoidal model. Under this scenario the standard F-statistic can no longer be assumed to follow an $F(m_{1}-m_{0},n-m_{1})$ distribution, but there is an asymptotic result (\citet{hamilton2020time}) based on Chisquare distribution with $m_{1}-m_{0}$ degrees of freedom. In Figure \ref{fig:spsamplesize} we compare the approximations of the CDF of the F-statistic for two different scenarios based on the sample size. For a relatively big sample size of 200 data points both the Saddlepoint approximation and the asymptotic result are close to the empirical CDF. When the sample size gets small however, as in Figure \ref{fig:spsamplesize} (Right) with 20 data points, the asymptotic result based on chi square no longer works well. The Saddlepoint on the other hand still approximates the distribution quite accurately.

\begin{figure}
		\centering
		\includegraphics[scale=0.45]{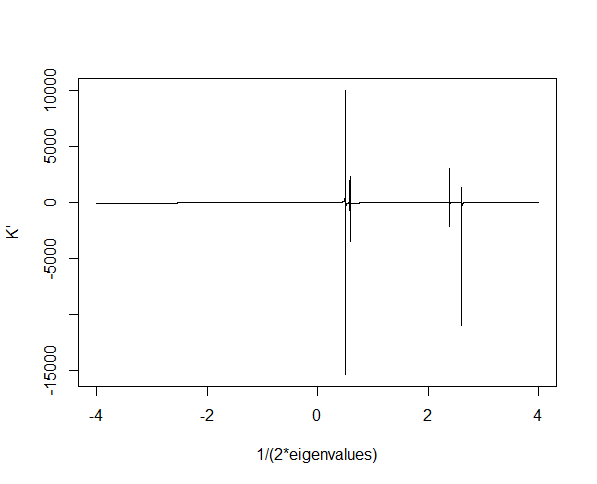}
		\includegraphics[scale=0.45]{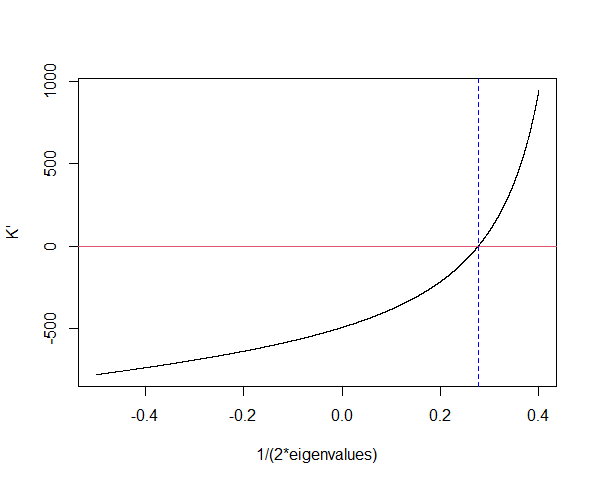}
		\caption{(Left) An example of the $\mathcal{K}'$ function, we see that its behaviour can be quite problematic making the choice of a suitable range to search for the solution important. (Right) For the same example only the range from the biggest negative to the smallest positive. The blue dashed line shows the solution of the Saddlepoint equation which in this case is 0.2763643.}
		\label{fig:saddle}
	\end{figure}

Finally, in Section \ref{s:impl} we discussed an optimal way in order to find a solution to the saddlepoint equation \eqref{spe}. An example of the general behaviour of $\mathcal{K}'$ can be seen in Figure \ref{fig:saddle}.

%%%%%%%%%%%%%%%%%%%%%%%%%%%%%%%%%%%%%%%%%%%%%%%%%

\subsection{Red noise vs White noise GPR model}\label{s:redvw}
\begin{figure}
		\centering
		\includegraphics[scale=0.45]{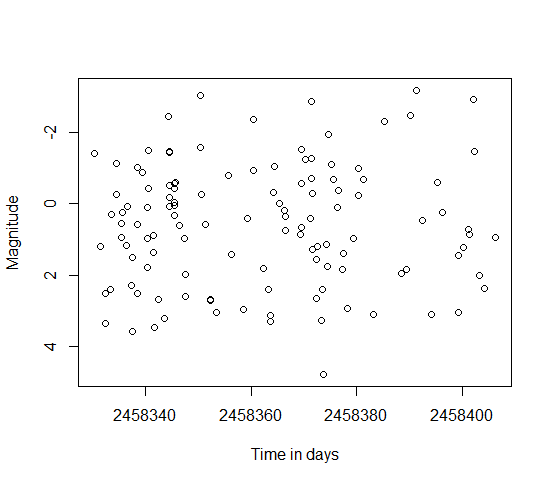}
		\includegraphics[scale=0.45]{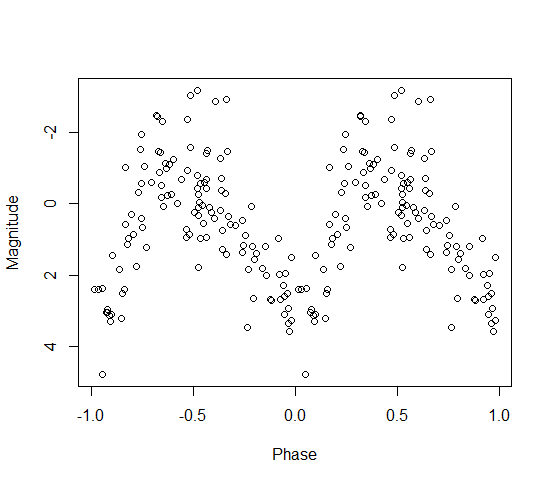}
		\caption{(Left) Simulated light curve from GPR prior with period at 5.1 days. The sampling is borrowed from a real light curve and this shape is used throughout the example of Table 1. (Right)he same simulated light curve folded into period of 5.1 days. }
		\label{fig:redex}
	\end{figure}

%\ak{why do we put this numerical example here while the saddle point example is in section 5?}
In Section \ref{s:hred} we saw how our test statistics $F$ and $CVF$ can be adjusted when our data are contaminated with red noise. Adjusting the models and tests to the correct noise assumption can improve drastically our results. More specifically, assuming a white noise model when the residuals are correlated can lead to periods failing to be identified and also to an increased number of falsely detected periods. As an example we will generate 100 light curves for 3 different correlated noise scenarios ($\rho=0.1, 0.5, 0.9$) with variance 1. The curves will be generated from a Gaussian process prior with a periodic kernel and period at 5.1 days. The mean behaviour of the curves will be fixed for all curves in all scenarios and it will be a realisation from the Gaussian process prior. We will borrow the sampling times from a real light curve (object 6785$\_$I from \citet{froebrich2021survey}). A typical example of the simulated data used here can be seen in Figure \ref{fig:redex}. Note that the performance of the methods is not affected by the magnitude's scale.\par 

%%%%%%%%%%%%%%%%%%%%%%%%%%%%%%%%%%%%%%%%%%%%%%%%%%%%

For these simulated light curves we will fit a Gaussian process regression model with a correlation structure $\C_{\rho}$ for AR(1) residuals. For this example, we estimate the parameters by minimizing the equivalent squared leave-one-out cross-validation error. 	
We will search for potential periods between 1 and 10 days with 1 decimal accuracy. Note that this particular period search grid is chosen for computational convenience. For real applications it is preferable to use the two-stage grid search method as seen in \citet{wang2012nonparametric} and \citet{reimann1994frequency}, whereby some N periods are initially chosen corresponding to the top N periodogram peaks based on a rough period grid, and then a finer period search grid is applied around each of the N chosen periods from the previous step. The results obtained by the red noise model will also be compared to those from the standard GPR model. We notice that for all three scenarios the red noise GPR model most of the time has the maximum peak of the F-statistic periodogram at the correct period, on the other hand, the white noise model (standard GPR) most of the time fails to do so. These results are summarized in Table 5 under the columns labelled as ``correct peak".
\begin{table}
\caption{A comparison between the red and white noise Gaussian process regression models fitted to red noise data. The columns labelled ``correct peak" show the percentage of the periodograms that had the maximum peak at correct period $p=5.1$. The columns labelled ``false periods" show the average number of periods falsely identified as valid.}
\begin{centering}
\begin{tabular}{lrrrr}
 \multicolumn{1}{c}{} &
      \multicolumn{2}{c}{GPR red noise}&\multicolumn{2}{c}{GPR white noise} \\ 
$\rho$ & \textbf{correct peak}& \textbf{false periods} &\textbf{correct peak}& \textbf{false periods}    \\
\hline
0.1& 90$\%$ & 1.34 &33$\%$ &3.56\\
0.5& 79$\%$ & 0.88 &25$\%$ &3.62\\
0.9& 84$\%$ & 1.20 &44$\%$ &2.90\\
\\

\end{tabular}\hfill
\end{centering}
\label{tab:rednoise}
\end{table}

Furthermore, we apply the generalized F-test as described in Section \ref{s:Hyp} to all simulated curves for a significance level $1-\alpha$ set to $0.95^{1/91}=0.9994365$.
This is the correction for multiple tests as seen in \citet{vsidak1967rectangular}, $91$ is the number of tests conducted which is the number of trial periods in our case. Almost always the correct period (at 5.1 days) was identified as significant (more than 99$\%$ of the time). When the wrong model (white noise) was used however, we had an increased number of false period discoveries compared to that of using the red noise GPR model. Specifically, after performing our period detection test, in many cases other periods, except for the true one, appeared to be falsely significant. These results are summarized in Table \ref{tab:rednoise} under the columns labelled as ``false periods", where we show the average number of periods falsely appearing to be significant, for each simulation scenario.
%%%%%%%%%%%%%%%%%%%%%%%%%%%%%%%%%%%
%\end{comment}
\subsection{Power comparison - Second example}
\begin{figure}[h]
		\centering
		\includegraphics[scale=0.46]{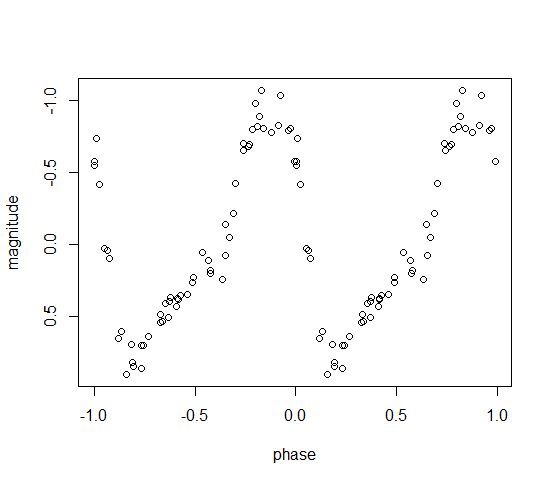}
		\includegraphics[scale=0.46]{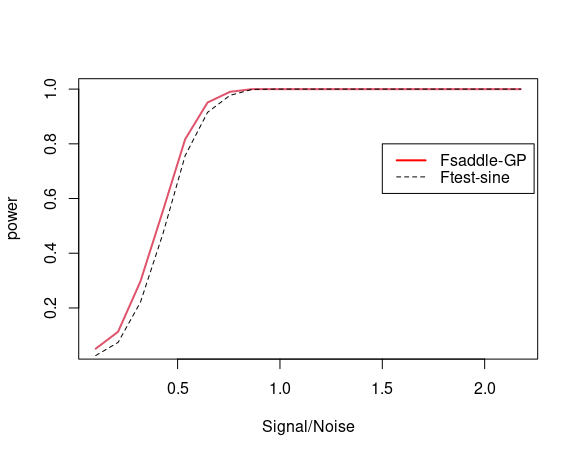}
		\caption{(Left) A phased light curve generated from the GPR model with period 5.2 days with SNR=6. This is a typical example of the shape of the light curves generated. (Right) The estimated power of the GPR generalized F-test of light curves generated for different SNR (red line). The dashed line shows the power of the standard F-test for a sinusoidal based fitting.}
		\label{fig:exampleimage}
	\end{figure}

Here we show another example similar to that shown in Section \ref{s:gprpow}. For this example we borrow the sampling of a randomly selected real light curve (object 3314$\_$I from \citet{froebrich2021survey}). We generate our periodic signal from a Gaussian prior using a periodic kernel as in \eqref{kernel} with period at 5.2 days. A typical example of the shape of our simulated data can be seen in Figure \ref{fig:exampleimage} (Left). We generate light curves for different signal to noise ratios, ranging from 0.01 to 6. The noise is generated from a zero-mean Normal distribution. We estimate the power for each different value of SNR based on 1000 repetitions. In Figure \ref{fig:exampleimage} (Right) we see the power of the generalized F-test at a $1-\alpha$ =99$\%$ significance level. We notice that the test performs very well with its power being estimated larger than 0.7 for even SNRs as small as 0.5. For a SNR larger or equal to 1 the power is constantly estimated close to 1. \par 
 We include in our plot the estimated power of the standard F-test using a sinusoidal model, for the same simulated data, as a reference. The tests in this example seem to perform similarly in terms of power with the generalized F-test being better. We performed the same analysis for the CVF-test too and obtained very similar results to those obtained from the generalized F-test. In Table 1 we can see a comparison of the power of generalized F and CVF tests for the same simulated data and SNR fixed to 0.65. The power is calculated for different significance levels. Both tests perform very well with their power estimated to be larger than 0.92 for a significance level of $1-\alpha=99.5\%$ for example. Of course the larger the $1-\alpha$ significance level the smaller the estimated power. We notice that the tests behave similarly with the CVF-test having a slightly larger power in this particular example. %\textbf{Finally we also calculated the probability of not rejecting the null hypothesis when the null is true by simulating purely noise data. In all cases the probability was 1 or really close to one for our proposed tests and the same holds for when a linear model and standard F-test is applied.}

%%%%%%%%%%%%%%%%%%%%%%%%%%%%%%%%%%%%%%%%%%%%%%%%%%%%

%%%%%%%%%%%%%%%%%%%%%%%%%%%%%%%%%%%%%%%%%%
\begin{table}
\caption{This table shows  the estimated power of the generalized F and CVF tests for different significance levels.}
\begin{centering}
\begin{tabular}{lrr}
 
\textbf{sig. level}&\textbf{$\mathbf{F}$ GPR }&\textbf{CVF GPR}\\
\hline
$0.95^{1/1}=0.95$& 0.987 &0.987  \\
$0.95^{1/10}=0.9948838$& 0.922 &0.924 \\
$0.95^{1/100}=0.9994872$& 0.714 &0.715\\
$0.95^{1/1000}=0.9999487$& 0.439 &0.446 
\\

\end{tabular}\hfill
\end{centering}
\label{tab:simGP}
\end{table}

%\end{document}

%\section{Appendix}
\end{document}